\documentclass[9pt,twocolumn,twoside]{opti_preprint} 

\usepackage{amsmath}
\newcommand\dd{\mathrm{d}}

\title{Hybrid photoacoustic/fluorescence microendoscopy through a multimode fiber using speckle illumination}

\author[1,*]{Antonio M. Caravaca-Aguirre$^{\dagger}$}
\author[2]{Sakshi Singh$^{\dagger}$}
\author[2]{Simon Labouesse}
\author[3] {Michael V. Baratta}
\author[2]{Rafael Piestun}
\author[1]{Emmanuel Bossy}

\affil[1]{Univ. Grenoble Alpes, LIPHY, F-38000 Grenoble, France}
\affil[2]{Department of Electrical, Computer, and Energy Engineering, University of Colorado, Boulder,Colorado 80309, USA}
\affil[3]{Department of Psychology and Neuroscience, University of Colorado Boulder, Boulder, CO, 80301, USA}

\affil[*]{Corresponding author: emmanuel.bossy@univ-grenoble-alpes.fr}

\begin{abstract}
We present an ultra-thin hybrid imaging system based on an optical multimode fiber (MMF) and an optical fiber hydrophone that combines optical resolution photoacoustic and fluorescence microscopy. To control the illumination at the distal tip of the MMF, a digital micromirror device modulates the amplitude of the optical wavefront which is coupled into the MMF. A set of pre-calibrated speckle illuminations combined with a reconstruction algorithm enables photoacoustic and fluorescence imaging of samples located at the distal tip of the fiber with optical resolution determined by the numerical aperture.

\end{abstract}

\begin{document}

\maketitle
\section{Introduction}
Imaging neural activity deep in the brain of small animals is limited by tissue scattering, and for a long time, it has only been possible to observe activity at the surface of the brain by observing the fluorescence signal of calcium indicators \cite{tian2009imaging}. Fluorescence imaging requires exogenous agents (dyes or genetically modified cells) to produce the contrast required to image. A complementary imaging modality that relies on endogenous contrast agents such as the light absorption is photoacoustic imaging. Photoacoustic imaging is an emerging multi-wave imaging modality that couples light excitation to acoustic detection via the photoacoustic effect (sound generation via light absorption). Lately, photoacoustic microscopy has been employed to obtain images of neural activity \cite{dean2016functional} with promising results. Both fluorescence and photoacoustic imaging offer unique advantages - while fluorescence imaging offers high sensitivity to specific molecular probes and the ability to look at multiple features simultaneously,  photoacoustic imaging provides specific sensitivity to non-radiative optical absorption. Combining the two modalities is therefore attractive and has proved effective for superficial tumor detection \cite{wang2010integrated,akers2010noninvasive,kim2010sentinel,maeda2015dual,zhang2017photoacoustic}.

\begin{figure*}[htbp]
\centering
\fbox{\includegraphics[width=\linewidth]{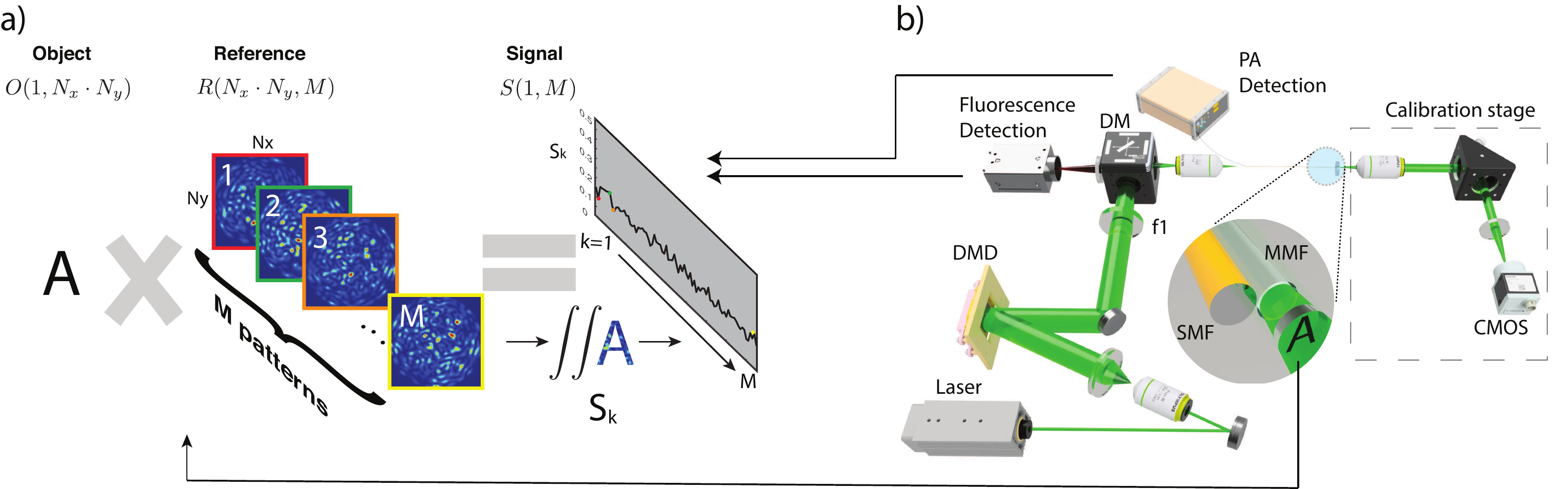}}
\caption{Principle of speckle illumination imaging through a MMF. A set of \textit{M} pre-calibrated speckle intensity patterns samples the object plane to generate \textit{M} corresponding $S_k$ signals proportional to the overlap between each speckle $R_k$ and the object. The integrated signal is detected using a single pixel detector. b) Sketch of the experimental setup used for both photoacoustic imaging (section~\ref{sec:results}.~\ref{subsec:PAresults}) and hybrid imaging (section~\ref{sec:results}.~\ref{subsec:Hybridresults}). }
\label{fig:Speckle_sketch}
\end{figure*}

For deep brain imaging with diffraction limited optical resolution, endoscopic approaches are required to avoid the loss of resolution due to scattering. At the cost of invasiveness, endoscopic approaches can both deliver light and collect signals  from deep regions. In fluorescence brain imaging, endoscopes are commonly based on fiber bundles or graded-index (GRIN) lenses \citep{oh2013optical} whose cross sections are in the order of a millimeter. Optical-resolution photoacoustic micro-endoscopes with a  similar footprint have been proposed using analogous approaches \citep{hajireza2011label,hajireza2013label, yang2015optical}. 
However, such cross sections are still larger than what is desirable to minimize neuronal tissue damage when the endoscope is inserted into the brain. Multimode fibers (MMF) are becoming a popular alternative employed to guide light and collect signals  thanks to  their small footprint as compared to fiber bundles with comparable imaging performances, and thanks to their efficient light collection for fluorescence imaging. Several groups have demonstrated the possibility to perform optical resolution imaging through  MMF~\cite{diLeonardo2011hologram,choi2012scanner, vcivzmar2012exploiting, bianchi2012multi, papadopoulos2013high, mahalati2013resolution, farahi2013dynamic, bianchi2013focusing, caravaca2013real, ploschner2015seeing, caravaca2017single, ohayon2018minimally, valley2016multimode,turtaev2018high}, including photoacoustic imaging~\cite{papadopoulos2013optical,simandoux2015optical,stasio2015towards}.

However, coherent light propagating through a MMF is seemingly randomized through fiber mode variations in phase velocity and potential mode-coupling, leading to a speckle pattern at the far end of the fiber (distal tip). Imaging through MMF therefore requires special methods that rely on a pre-calibration of the fiber to determine its input-output relation, namely the transmission matrix (TM) \cite{bianchi2012multi,vcivzmar2012exploiting, papadopoulos2013high, caravaca2013real, bianchi2013focusing, ploschner2015seeing}. Hence, given the fiber TM, we can control the illuminations used to sample the output plane by modifying the wavefront of the input illuminations, and collect a feedback signal such as a fluorescence or photoacoustic emission, to recover the object. A common choice of such controlled illuminations is scanning focal spots. However, this local sampling approach needs scanning of \textit{N} focal spots to obtain an \textit{N} pixel object reconstruction. Moreover, generating focal spots at the distal end requires complete calibration of the fiber complex TM. It is worth noticing that the fiber needs to stay mostly unperturbed during the experiment; otherwise the TM changes, reducing the ability to control the intensity distribution at the distal tip \cite{caravaca2017single}. 

Here we employ an approach that does not require focusing, based on sampling the output plane using the natural output speckle patterns of a stationary MMF. In addition to being readily produced by propagation through MMF, speckle patterns have also been shown to be ideal for compressive sensing \cite{valley2016multimode, wang2018ultrafast}. This idea was first proposed by Bolshtyansky et al. \cite{bolshtyansky1996transmission} who simulated the total integrated signal coming from a reflective object illuminated with speckles produced by a MMF and demonstrated coherent imaging. The concept was later  demonstrated experimentally through MMF first with reflective samples\cite{mahalati2013resolution}, and more recently with fluorescence beads\cite{amitonova2018compressive,Labouesse2018random}. In photoacoustic imaging, the same approach was also implemented for imaging test samples through a scattering  sample \cite{poisson2017multiple} and through MMFs \cite{caravaca2018speckle}. Although this approach still requires a calibration step, it is simpler because it only requires to measure the optical speckle intensities as opposed to wavefront shaping based methods that require speckle field measurements. Additionally the nonlocal sampling has shown better compression ratios in noisy environments than local approaches \cite{akhlaghi2015compressive} and is advantageous in imaging sparse samples that are often encountered in biological imaging.

In this work, we show that the above concept can also be extended to implement both photoacoustic and fluorescence endoscopic imaging through a MMF. Combining the pre-recorded speckle illuminations and the corresponding fluorescence and photoacoustic signals from the object at the distal tip of the MMF with reconstruction algorithms, we obtain images of biological test samples in vitro with both modalities with a minimally invasive microendoscope. To our knowledge, this is the first demonstration of an ultra-thin MMF imaging system capable of optical resolution photoacoustic imaging and fluorescence imaging at the same time with micrometer resolution.

\section{Methods}
\label{sec:Methods}
\subsection{Principle of imaging using speckle illumination} 

Figure \ref{fig:Speckle_sketch} a illustrates the principle of the measurements for imaging with speckle illumination. A set of $M$ random patterns is projected on the fiber input (proximal tip), generating different speckle illuminations $R_k(x,y)$ at the fiber output (distal tip). This reference set is  first measured using a camera before the object is placed at the distal tip, as a calibration step. Each output pattern has a different distribution of speckle grains and therefore probes the field-of-view differently from other patterns.

During the measurement step, the object $O(x,y)$ to be reconstructed is illuminated by the same set of $M$ speckle patterns and a feedback signal from the whole field-of-view is captured through the same fiber by a single-pixel detector.In our  work, the single-pixel detector is either a fluorescence detector or an acoustic detector. For both photoacoustic and fluorescence measurement, the signal, $S_k$, received by a single pixel detector when the object is illuminated by each speckle pattern is modeled as the overlap integral of the object and the illumination pattern:

\begin{equation}
S_k = \iint_{A}{R_{k}(x,y)O(x,y)}\dd x \dd y.
\label{eq:signal}
\end{equation}

The intensity fluctuations from speckle pattern to speckle pattern in the object plane translates into fluctuations of the signal $S_k$, thereby encoding sample information at the positions at which the speckle grains overlap with the object, reminiscent of the working principle of a single pixel camera \cite{duarte2008single}.

\subsection{Image reconstruction with a sparsity-constrained optimization}

From the set of $M$ measurement values, there are various  ways to reconstruct an estimate of the object~\cite{akhlaghi2015compressive,poisson2017multiple}, including correlation-based image reconstruction or resolution of a linear problem, possibly incorporating prior information on the object~\cite{katz2009compressive,shin2017compressive,guerit2018compressive} . A comparison of the results obtained from our experimental measurement with three different approaches is provided in the supplementary information. In our work, the objects of interest are sparse in the real space, and the best results were obtained with a sparsity-constrained solution of a linear problem formulated in a matrix formalism, following the approach introduced for ghost imaging \cite{katz2009compressive}.

In this approach, the discretization of equation \ref{eq:signal} over the N pixels of the camera yields a linear problem in a matrix form written as 
\begin{equation}
\mathbf{S}=\mathbf{R}\times \mathbf{O}.
\label{eq:matrix}
\end{equation}
In equation \ref{eq:matrix}, \textbf{S} is a $M \times 1$ vector which represents the measurements, \textbf{R} is an $M \times N$ matrix of the set of M illumination patterns measured on the camera during the calibration step, and \textbf{O} now represents a discretized version of the object as a $N \times 1$ vector. An estimate of the object \textbf{O} can be computed by solving the following minimization problem \cite{katz2009compressive}:
\begin{equation}
\min_O||\mathbf{R}\times \mathbf{O}-\mathbf{S}||^2_2+\gamma||\mathbf{O}||_1.
\label{eq:cost_function}
\end{equation}
The first term of the cost function ensures that the reconstructed object fits the data while the regularization term with a L1-norm favors sparse objects. $\gamma$ is a regularization parameter that needs to be tuned depending on the signal-to-noise ratio and the degree of sparsity in the object. Such formulation allows to find an estimate of the object \textbf{O} even when the matrix \textbf{R} is non invertible (either because $N>M$ or because the $M$ speckle patterns may not form a basis due to correlations). The regularization term is needed to account for noise in the measurements and the L1-norm favors sparse estimates. The resolution of equation \ref{eq:cost_function} was performed numerically with a Fast Iterative Shrinkage-Thresholding Algorithm (FISTA), further described in the Supplementary Material.

\subsection{Speckle illumination pattern selection}
The quality of image recovery is  significantly dependent on two properties of the set of references speckle patterns. First is the sampling efficiency: the ability to perfectly recover an object depends on the completeness of the sampling of the object plane by the output illumination set.  An ideal set of illuminations would form a complete basis set which can encode information about every point at the fiber distal tip. In practice, generating such an illumination set is limited by the efficiency of exciting all the modes of the fiber, and the number of available speckle patterns can thus be less than the number of pixel to reconstruct. The second property is compressibility. Since speed is an important factor in imaging through dynamic scattering media, recovering an object using a number of illuminations as small as possible is highly desirable, and made possible by the proposed approached when sparse objects are concerned \cite{katz2009compressive}.

To maximize the compression ratio $N/M$, while minimally affecting sampling efficiency, the correlations between individual illuminations must be minimized so that each of them is able to retrieve unique information about the sample. Those correlations can be quantified by measuring their mutual coherence, defined as
\begin{equation}
\mu_{ij}=\frac{R_i \cdot{} R_j}{|R_i|\cdot{}|R_j|}
\label{normalization}
\end{equation}
where $\mu_{ij}$ is the mutual coherence between the i\textsuperscript{th} and j\textsuperscript{th} illuminations. The $M \times M$ mutual coherence matrix can be constructed containing the correlation of each speckle with every other speckle. 

The optimization to select the speckles with reduced correlation involves the following steps:
\begin{enumerate}
\item{Set the speckle self-correlation terms on the diagonal elements of the coherence matrix, to zero. This way only correlations between different speckles are analyzed.}
\item{Calculate the norm of each row in the mutual coherence matrix. The row $R_i$ with maximum norm signifies that the i\textsuperscript{th} speckle has maximum correlations with all other speckles in the illumination set.}
\item{Set the $i\textsuperscript{th} $ row and column in the coherence matrix to zero and note the index, i, of the speckle to be deleted from the illumination set. The new coherence matrix corresponds to correlations among M-1 illuminations.}
\item{Repeat 2 and 3 till the number of non-zero rows in the coherence matrix becomes equal to the size of the desired optimized illumination set.}
\end{enumerate}
Note that the coherence matrix is symmetric, which means the above optimization can equivalently be performed on columns instead of rows of the coherence matrix, leading to the same results.

\subsection{Experimental setup}

Three types of proof-of-principle experiments were performed in this work, namely photoacoustic microscopy alone, fluorescence microscopy alone, and hybrid photoacoustic/fluorescence microscopy. 

A schematic description of the experimental setup used for both photoacoustic microscopy and the hybrid photoacoustic/fluorescence microscopy (see results in sections \ref{sec:results}.\ref{subsec:PAresults} and \ref{sec:results}.\ref{subsec:Hybridresults}) is shown on Figure \ref{fig:Speckle_sketch} b.  The excitation light is provided by a a pulsed laser (Cobolt Tor\texttrademark  series, 3ns pulse duration, 532nm, 7kHz repetition rate). A digital micromirror device (DMD, Vialux \textregistered{}, model V-7001) is used to modulate the optical wavefront at a repetition rate of 22 KHz. The laser beam is first expanded to match the dimensions of the DMD, and the DMD is then imaged with a 4f-system onto the input facet of a multimode fiber (MMF). A 8-cm long MMF (Thorlabs DCF13) is used to guide light to the object plane and collect fluorescence. The outer diameter of the MMF (without protective cladding) is 125 $\mu$m. The 105 $\mu$m-diameter multimode core is used to guide the source light, and also collect fluorescence. The fluorescence signal is measured with a photomultiplier tube (PMT, Hamamatsu, model H7422) placed behind a dichroic mirror. For photoacoustic measurements, a fiber-optic hydrophone (FOH, Precision Acoustics \textregistered{}), with an outer diameter of 125 $\mu m$ (without protective cladding) is attached parallel to the MMF. Both fibers are held together inside a metallic cannula to make sure they remain attached and avoid relative motion. A few millimeters of the distal tip of both fibers stick out of the metallic cannula (as shown in the inset of Figure \ref{fig:Speckle_sketch}.b), as an imaging head with a minimal total footprint of about 250 $\mu m$. 

For the calibration step, a set of $M$ random patterns (M on the order of a few thousands) with a 50\% fill factor are projected onto the DMD producing a set of speckle patterns at the distal tip of the MMF. The speckle patterns are measured with a CMOS camera (Basler \textregistered{} acA1920-155um) that, after a 20x magnification, images a plane located $\sim 50\mu$m away from the MMF distal. After the calibration is performed, the sample is placed in the same imaging plane and the same set of speckle patterns are projected onto the sample. In the sample plane, the image pixel size was 0.36$\mu$m.

For each speckle pattern, the setup allows to detect and  record either a photoacoustic signal with the FOH or a fluorescence signal with the PMT. Both detectors act as single pixel detectors collecting a signal $S_k$ integrating information from the whole field-of-view. The time-resolved voltage signals from the detectors are digitized and recorded with a data acquisition card (Gage Razor Express 1622). Finally, the recorded temporal traces are processed to obtain a scalar value $S_k$ for each type of signal (photoacoustic or fluorescence).  For photoacoustic measurements, $S_k$ is defined as the area under the envelope of the FOH pulsed signal. For fluorescence measurements, $S_k$ is defined as the peak value of the PMT pulsed signal. To improve the signal-to-noise ratio (SNR) when needed, signals were averaged over repeated measurements for each speckle pattern, as specified further below in Sec. \ref{sec:results}.The synchronization of all the parts of the system is controlled by a BNC pulse generator.

The fluorescence results described in section \ref{sec:results}.\ref{subsec:Fluoresults} were obtained with a slightly different setup for practical reasons related to samples and material availability in the two different facilities involved in this work. The corresponding experimental setup is described in detail in \cite{caravaca2017single,ohayon2018minimally}, and differs from the setup above in particular by the fact that a continuous laser was used for the experiments. Additionnaly, an EMCCD camera
(Andor iXon+) was used to collect fluorescence from the input/proximal tip of the MMF. The measurement methodology was however strictly identical with both setups.
\section{Experimental Results}

\label{sec:results}

Three types of proof-of-principle experiments were performed in our work. In section \ref{sec:results}.\ref{subsec:PAresults} we  present images obtained by photoacoustic microendoscopy  of various types of samples, including \textit{in vitro} red blood cells. In section \ref{sec:results}.\ref{subsec:Fluoresults} we  present images obtained by fluorescence microendoscopy  of various types of samples, including \textit{in vitro}  images of fluorescent beads in a slice of mouse brain. We also illustrate how the set of illumination patterns can be chosen to optimize the quality of the reconstructed image. Finally, we show in section \ref{sec:results}.\ref{subsec:Hybridresults} how both imaging modalities can be combined into a unique optical setup to experimentally validate the compressive hybrid imaging of red blood cells and fluorescent particles simultaneously.

\subsection{Photoacoustic imaging}
\label{subsec:PAresults}
\begin{figure}[htbp]
\centering
\fbox{\includegraphics[width=\linewidth]{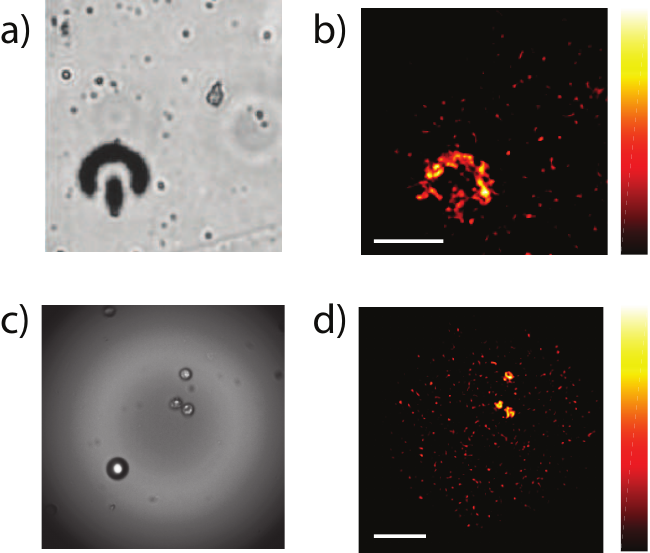}}
\caption{Experimental results for photoacoustic microendoscopy.  a) Bright-field microscope image of an absorbing micro-structure. b) Reconstructed 200x200 pixels (M=40,000) photoacoustic image, obtained with $M=4096$ speckle patterns. c) Bright-field microscope image showing three red blood cells and a fluorescent bead d) Reconstructed 300x300 pixels (M=90,000) photoacoustic image, obtained with with $M=4096$ speckle patterns. The scale bar is 30 $\mu$m.}

\label{fig:Results_PA}
\end{figure}

In this part, for each pattern projected at the DMD, the photoacoustic signal was averaged between 100 and 1000 laser pulses to obtain a sufficient SNR. Two types of samples were used to illustrate the photoacoustic imaging capability of the setup shown in Fig. \ref{fig:Speckle_sketch} b. As a first, well-controlled test sample, we used an absorbing micro-structure photoplotted on a polymer film (Selba S.A, Versoix, Switzerland), shown on Fig. \ref{fig:Results_PA} a ("power-on" logo). The field of view corresponds to a 200x200 pixels scene, corresponding to $N=40,000$. For each pattern projected at the DMD, the photoacoustic signal was averaged over 100 laser pulses to obtain a sufficient SNR. For the reconstructed photoacoustic image shown in Fig. \ref{fig:Results_PA} b, only $M=4096$ random speckle patterns were used to compute the photoacoustic values $S_k$, and both parts of the "power on" sign are imaged with a very low background signal. It is important to notice that the "granular" appearance is due to the relatively low number of speckle patterns used (about $10\%$ of the total number of reconstructed pixels). This artifact can be reduced by increasing the number of patterns or repeating the experiment with different speckle realizations optimized using the method described in \ref{sec:Methods} C. To further demonstrate the performance of the system on a more relevant biological sample, we used the same system to obtain photoacoustic images of red blood cells previously washed and deposited on phosphate buffered saline (PBS). Fig. \ref{fig:Results_PA} c presents a bright field microscope image of the sample, which shows three red blood cells and a Nile red fluorescent particle (big circle on the left bottom) added to the buffer to demonstrate that the fluorescent dye does not produce any detectable photoacoustic signal. $M=4096$ speckle patterns were used to reconstruct the 300x300 pixels photoacoustic image ($N=90,000$) shown in Fig. \ref{fig:Results_PA} d: the three red blood cells are clearly visible and well resolved on the reconstructed image, and the fluorescent bead is invisible as expected.

\begin{figure}[htbp]
\centering
\fbox{\includegraphics[width=\linewidth]{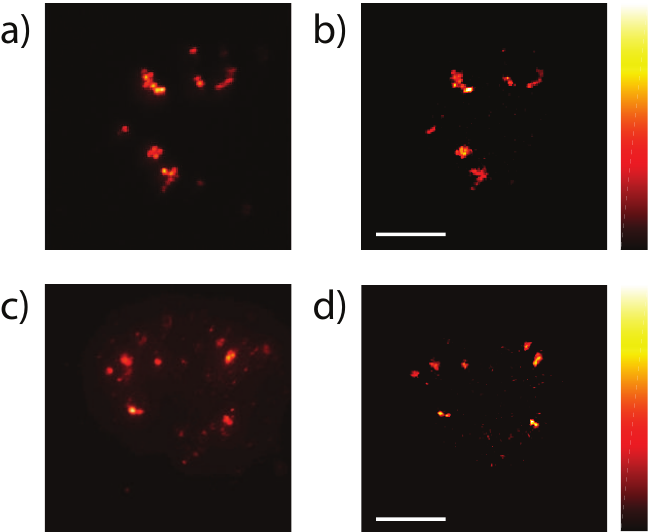}}
\caption{Experimental results of fluorescence imaging. a,b) Imaging of 4 $\mu$m orange beads from TetraSpeck Fluorescent Microspheres Sampler kit: a) Widefield fluorescence image captured on the distal side of the fiber after averaging over 4000 speckles. b) 192x192 pixel object reconstruction using 10000 speckle illuminations. (c,d) Fluorescence imaging of red fluorescent retrobeads in a mouse brain slice: c) Widefield fluorescence image captured as described for Fig. 3(a) d) 192x192 pixel object reconstruction using 10000 speckle patterns.}
\label{fig:Results_Fluo}
\end{figure}

\subsection{Fluorescence imaging}
\label{subsec:Fluoresults}
Two types of samples containing fluorescent beads were imaged with a similar experimental setup described in the supplementary information. In all imaging experiments through the fiber, i.e.  when fluorescent light was collected from the proximal tip of the fiber after propagation through the MMF, the raw fluorescence data contained contribution from the autofluorescence of the fiber, more prominently from the fiber cladding. This autofluorescence from the fiber cladding was discarded (by operating in the image acquisition mode of the EMCCD) to selectively detect the fluorescence signal from the sample guided through the fiber core. Additionally, a background signal in the absence of the object was subtracted from all the data before running the reconstruction algorithm.

Fig. \ref{fig:Results_Fluo} a shows a reference widefield fluorescent image of 4 $\mu$m orange beads from a  TetraSpeck Fluorescent Microspheres Sampler kit. This reference fluorescent image was obtained with a CMOS camera directly imaging the fluorescent sources at the output/distal side, by averaging the fluorescence images corresponding to 4000 different speckles patterns. The 192x192 pixels ($N=36864$) image from fluorescence collected at the input/proximal side of  fiber is shown in Fig. \ref{fig:Results_Fluo} b: this reconstructed image demonstrates that the complex distribution of beads is well-recovered while preserving the boundaries of both individual and clustered beads. $M=10,000$ speckle illuminations were used to reconstruct the image, corresponding to about $25\%$ of the total number of reconstructed pixels.

We also performed imaging of red fluorecent retrobeads (0.05 - 0.2$\mu$m) from Lumafluor microinjected into the dorsomedial striatum (DMS) of a mouse brain, which was then sliced and mounted on a microscope slide. As for the first sample, Fig. \ref{fig:Results_Fluo} c shows a reference widefield fluorescent image of the sample. Fig. \ref{fig:Results_Fluo} shows the corresponding image reconstructed with our approach, and also clearly demonstrates the recovery of individual clusters of retrobeads in neurons.
It can be observed that the resolution of the reconstructed object is dictated by the grain size of the speckle produced by the excitation wavelength, thanks to which the reconstruction image on Fig. \ref{fig:Results_Fluo} d is better resolved than the reference widefield fluorescence image on Fig. \ref{fig:Results_Fluo} c obtained by direct imaging with the fluorescent light. Moreover, the sparsity assumption  allows a better z-sectioning by eliminating the out-of-focus features from the reconstruction image seen in the reference widefield fluorescence image on Fig. \ref{fig:Results_Fluo} c.

\begin{figure*}[htbp]
\centering
\fbox{\includegraphics[width=\linewidth]{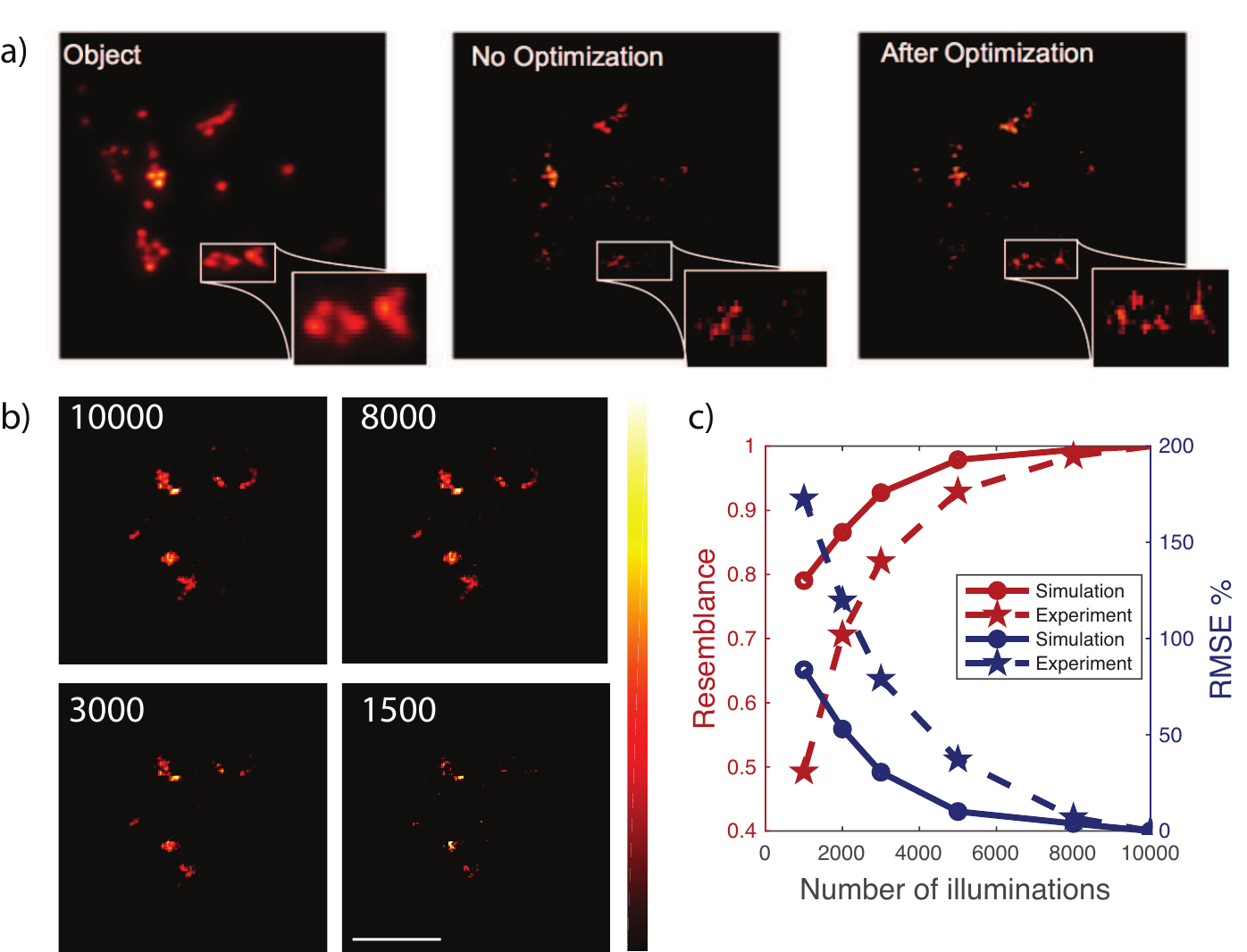}}
\caption{Influence of the set of speckle patterns on the reconstructed images of of 4 $\mu$m orange beads from TetraSpeck Fluorescent Microspheres Sampler kit using optimized illuminations. (a) Left: widefield fluoresence image. Middle: image reconstructed with $M=3000$ speckle patterns randomly chosen out of 10000 speckles patterns. Right: image reconstructed with $M=3000$ speckle patterns chosen as to minimize cross-correlations between speckle patterns. The image obtained without optimizing the choice of speckle patterns misses some features of the original object. The image obtained after optimization illustrates the  improved sampling efficiency for an optimized choice of the speckle patterns. (b) Reconstructed images using different number $M$ of speckle patterns. The set of 8000, 3000 and 1500 speckle patterns were chosen optimally out of the 10000 available speckle patterns in order to minimize the cross-correlations inside each speckle set, as for (a). (c) Plots of resemblance and relative mean square error (RMSE) between reconstructed images and the reference image ($M_{\mathrm{max}}=10,000$), as a function of the size $M$ of the speckle set.}

\label{fig:Compressed_sensing}
\end{figure*}

We further used the measurement values obtained with the TetraSpeck microscospheres to study the influence of the set of speckle patterns used to perform the measurements on the image quality. Fig. \ref{fig:Compressed_sensing} a illustrates how chosing $M=3000$ speckles patterns influences the quality of the reconstructed image: if the 3000 speckle patterns are chosen randomly out of a larger set of 10000 patterns, the reconstructed image is shown in Fig. \ref{fig:Compressed_sensing} a (middle image, "No Optimization"). If the 3000 speckle patterns are chosen such as to minimize cross-correlation between speckle patterns as presented in Sec. \ref{sec:Methods}, a significantly better image can be obtained as illustrated in Fig. \ref{fig:Compressed_sensing} a (right image, "After Optimization").

In Fig. \ref{fig:Compressed_sensing} b and c, we further illustrate how the size of the speckle pattern set influences the quality of the reconstructed image. We take a set with $M=10,000$ to define a reference reconstructed image, and study the quality of the images reconstructed with optimized subsets of smaller size. Fig. \ref{fig:Compressed_sensing} b shows that for a speckle set of size as low as $M=3000$,  a $N=40,000$ pixel image remains recoverable while maintaining a good qualitative resemblance with the reference image, whereas for $M=1500$, features of the object are obviously lost.  Fig. \ref{fig:Compressed_sensing} c provides a more quantitative insight into how the size of the speckle set influences the reconstruction:  Fig.\ref{fig:Compressed_sensing} c plots the relative mean square error as well as the resemblance (estimated by cross-correlation) of the reconstruction images as a function of M with respect to the reference image. Simulations data are obtained by using images reconstructed from signal values computed from equation \ref{eq:matrix} with R containing each set of experimentally measured speckle patterns and the object O as the image \ref{fig:Results_Fluo} b with M=10,000. Both simulated and experimental curves show a decay in error and increase in resemblance with increasing number of illuminations, as expected. The effect is more pronounced in the case of the experimental data, which is likely due to the influence of experimental noise. 

\subsection{Hybrid imaging}
\label{subsec:Hybridresults}

\begin{figure}[htbp]
\centering
\fbox{\includegraphics[width=\linewidth]{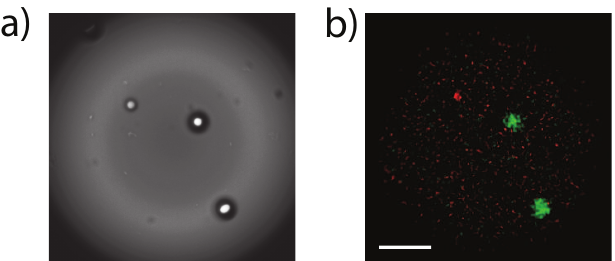}}
\caption{Hybrid imaging of red blood cells and 11 $\mu m$ diameter fluorescence particles dyed with nile red. a) Bright-field microscope image of the sample at the distal tip using the calibration CMOS camera b) False color hybrid image reconstruction of the red blood cell (in red) and the fluorescence particles (in green). Scale bar is 30 $\mu$m.}
\label{fig:Results_hybrid}
\end{figure}

To demonstrate the hybrid imaging capability of our system, we finally performed an experiment where we obtain simultaneously the photoacoustic and fluorescence image of the same sample. The system used is shown in Fig. \ref{fig:Speckle_sketch} b. We used a diluted solution composed of red blood cells and 11 $\mu$m diameter fluorescent particles (Nile red) in PBS. Figure \ref{fig:Results_hybrid} a shows the widefield microscope image of the sample using incoherent illumination through the MMF where two fluorescence particles and one red blood cell can be identified. To obtain both the absorption and fluorescence images, two different illumination configurations had to be set: fluorescence imaging requires a low energy to avoid photo-bleaching of the fluorophores, while photoacoustic imaging requires high energy to have a sufficient signal-to-noise ratio. Therefore, we first performed the fluorescence measurement projecting all the speckle patterns with low energy ($\sim 4$nJ per pulse at the distal tip) and then the photoacoustic detection by projecting again the same DMD patterns at high energy ($\sim 4\mu$J per pulse at the distal tip).

Figure \ref{fig:Results_hybrid} b shows the false-color reconstructed photoacoustic image (red) and fluorescence image (green) from the set of signals corresponding to 2048 speckle patterns. The area calibrated with the CMOS camera is 300x300 pixels, corresponding to 150 $\mu m^2$. The photoacoustic image clearly shows the ability to resolve single red blood cells and it does not have any spurious signal coming from the fluorescence particles as expected. In the fluorescence case, the signal recorded by the PMT only has contributions from the fluorescence particles.

\section{Discussion and conclusions}

In conclusion, we have presented an ultra-thin system capable of performing simultaneous florescence and photoacoustic microscopy. The footprint of the probe head is 250$\mu m$ x 125 $\mu$m, which makes it the thinest device capable of hybrid imaging to our knowledge. The reference-free calibration presented here, demonstrating the ability to recover the object using intensity-only measurements, simplifies the optical system compared to other approaches based on learning the transmission matrix of the MMF. Furthermore, we presented a way of optimizing the illumination set for a stationary MMF system, such that maximum information about the object can be recovered using the smallest possible number of illuminations, hence further boosting the imaging speed. 
Overall, the combination of photoacoustic and fluorescence imaging - two prevalent imaging modalities for in-vivo imaging through biological tissue - put together in a 250 $\mu m$ thin fiber system makes for a powerful tool that could have a great impact in a range of biomedical applications, and specifically for deep brain neural activity detection.

\section*{Funding Information}
This project has received funding from the European Research Council (ERC) under the European  Union’s Horizon 2020 research and innovation programme  (project COHERENCE, grant agreement 681514), from the People Program (Marie Curie Actions) of the European Union’s Seventh Framework Program (FP7/2007-2013) under REA grant agreement PCOFUND-GA-2013-609102, through the PRESTIGE program coordinated by Campus France and the Marie Skłodowska-Curie Individual Fellowship (project DARWIN 750420), the National Institute of Health (award REY026436A) and from NSF (award 1611513).

\section*{Acknowledgments}
We acknowledge the efforts of Sylvie Costrel for the red blood cells sample preparation. 

\section*{Supplemental Documents}

\bigskip \noindent See \href{link}{Supplement Material } for supporting content.

\bigskip \noindent $^{\dagger}$ This authors contributed equally to this work
\bibliography{sample}

\clearpage
\onecolumn
\begin{center}
\textbf{\Huge Hybrid photoacoustic/fluorescence microendoscopy through a multimode fiber using speckle illumination: supplementary material}
\end{center}
\setcounter{equation}{0}
\setcounter{section}{0}
\setcounter{figure}{0}
\setcounter{table}{0}
\makeatletter
\renewcommand{\thesection}{S\arabic{section}}
\renewcommand{\theequation}{S\arabic{equation}}
\renewcommand{\thefigure}{S\arabic{figure}}

\section*{}
This  document  provides  supplementary  information  to "Hybrid photoacoustic/fluorescence microendoscopy through a multimode fiber using speckle illumination". We show a comparison between three different image reconstruction algorithms and we explain the details of the experimental setup used for fluorescence imaging.

\section{Image reconstruction algorithms}

\label{Reconstruction_algorithms}
\begin{figure}[htbp]
\centering
\fbox{\includegraphics[width=\linewidth]{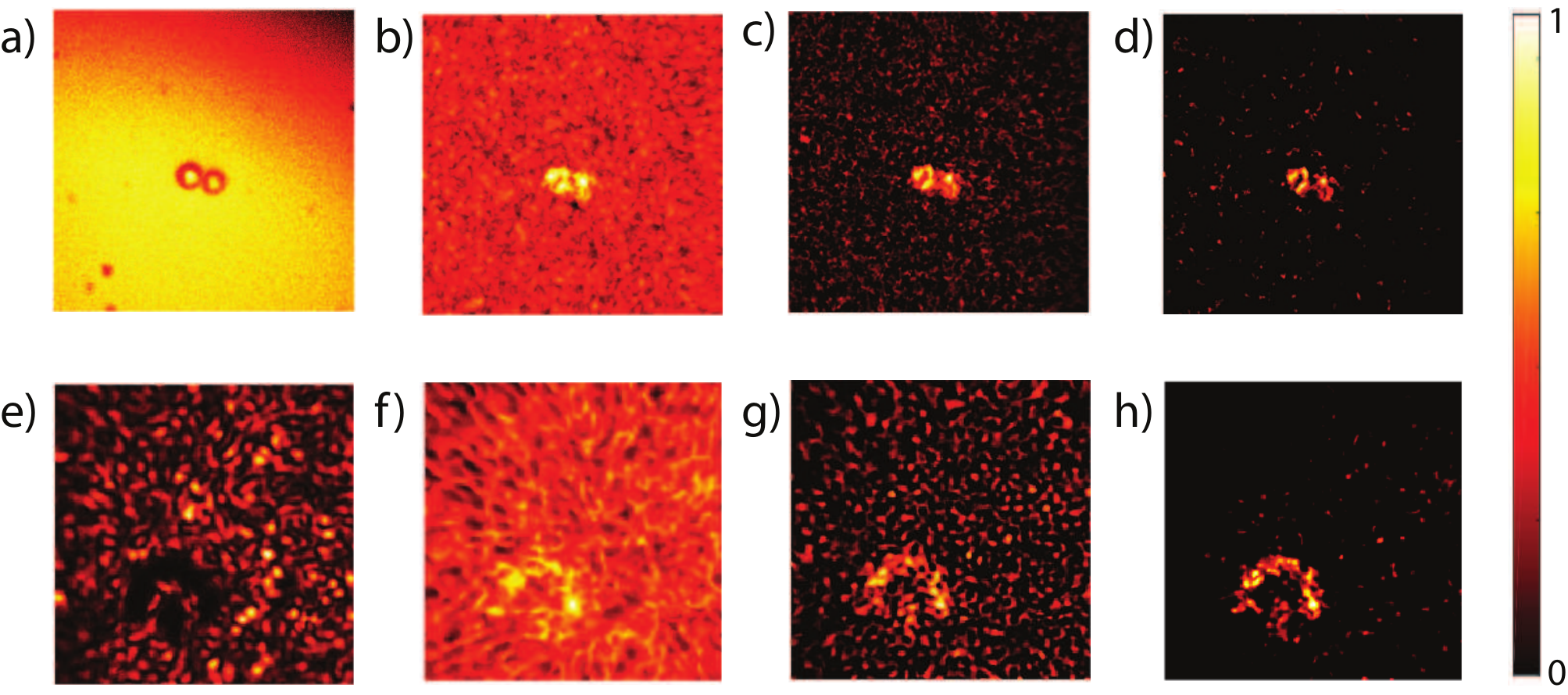}}
\caption{Comparison of different reconstruction algorithms. Fluorescence imaging of (a-d) two 11$\mu$m diameter fluorescence particles and photoacoustic imaging of(e-h) an absorbing "power on" sign. a,e) Widefield microscopy image. Reconstruction images using (b,f) the correlation based method (c,g) the minimization optimization without regularization term and (d,h) with regularization term and positivity constrain. }
\label{fig:MethodComparison}
\end{figure}

There are several methods to reconstruct the object as mentioned in Section 2. One approach based on simple correlations, as used in \cite{akhlaghi2015compressive}, is based on the following estimation of the object:
\begin{equation}
O(x,y)=\frac{1}{M}\sum ^M _{r=1} (S_k-\langle S \rangle) R_k(x,y),
\end{equation}
In this approach, pixel-wise cross-correlations between the signals value $S_k$ and the calibrated intensity values directly provide an estimate of the object. As introduced in the main text, the object can also be estimated through a linear problem formulation by minimizing a cost-function. One simple possible approach in this case is to solve the following problem
\begin{equation}
\min_O||\mathbf{R}\times \mathbf{O}-\mathbf{S}||^2_2,
\end{equation}
by using the Moore-Penrose pseudo-inverse "$R^{-1}$" of R and computing $O="R^{-1}"\times S$. This methods is however known to be sensitive to measurement noise, and the approach described in the main text was to minimize a cost function with a L1-based regularization term and positivity constrain as
\begin{equation}
\min_O||\mathbf{R}\times \mathbf{O}-\mathbf{S}||^2_2+\gamma||\mathbf{O}||_1, O\geq 0
\end{equation}
The L1 penalty term is known to favor sparse objects. 
To solve the above minimization problem, we use Fast Iterative Shrinkage-Thresholding Algorithm (FISTA) \cite{beck2009fast} a state of the art fast algorithm to find a solution of our optimization problem:

\begin{equation}
y^k=max(2 \mathbf{A}^\mathbf{t}(A \times x^\mathbf{k+1}-\mathbf{S}),0)
\end{equation}
\begin{equation}
\mathbf{t}^\mathbf{k+1}=\frac{1+\sqrt{1+4\mathbf{t}^\mathbf{k^2}}}{2}
\end{equation}
\begin{equation}
\mathbf{x}^\mathbf{k+1}=y^\mathbf{k}+(\frac{t^\mathbf{k}-1}{t^\mathbf{k+1}})\times (\mathbf{y}^\mathbf{k}-\mathbf{y}^\mathbf{k-1})
\end{equation}
The convergence of FISTA is theoretically guaranteed with a convergence rate of $O(1/k^2 )$. 
Figure \ref{fig:MethodComparison} illustrates the performances of the three approaches described above for both fluorescence and photoacoustic imaging. The sparsity- and positivity-constrained reconstruction is cleary superior to the two other approaches, and was therefore used for all the reconstruction presented in the main manuscript.

\section{Experimental setup for Fluorescence imaging}
The fluorescence imaging setup used for experiments described in section 3B of the main paper consists of a 532 nm CW Coherent Verdi-G laser which passes through a spatial filter SF and is incident on a  TI-DLP Discovery 4100 digital micro-mirror device (DMD). The DMD performs phase modulation on the incident wavefront using off-axis computer generated holography, as described in \cite{caravaca2017single}. The incident wavefront is modulated using 9216 independent input modes and imaged onto the back focal plane of a 20x microscope objective MO1 which then couples it into the fiber. The distal end of the fiber is imaged using a 40x microscope objective MO2 and lens L3 onto a CMOS camera (Hamamatsu Orca Flash 2.8). For calibration, we project patterns from the binary random orthonormal basis set  on the objective back focal plane to produce different speckles at the fiber distal end. After calibration, we mount the fluorescent sample on a sample holder SH at the fiber distal end and project the same set of patterns used for calibration.  For each projection, we record the corresponding fluorescence signal coming back to the input end of the fiber using an Andor iXon+ Electron multiplying gain CCD (EMCCD). The excitation photons are rejected using a dichroic mirror DM (Chroma ZT532rdc-UF1) and a bandpass filter (Chroma ET590/50m). 
\begin{figure}[htbp]
\centering
\includegraphics[width=\linewidth]{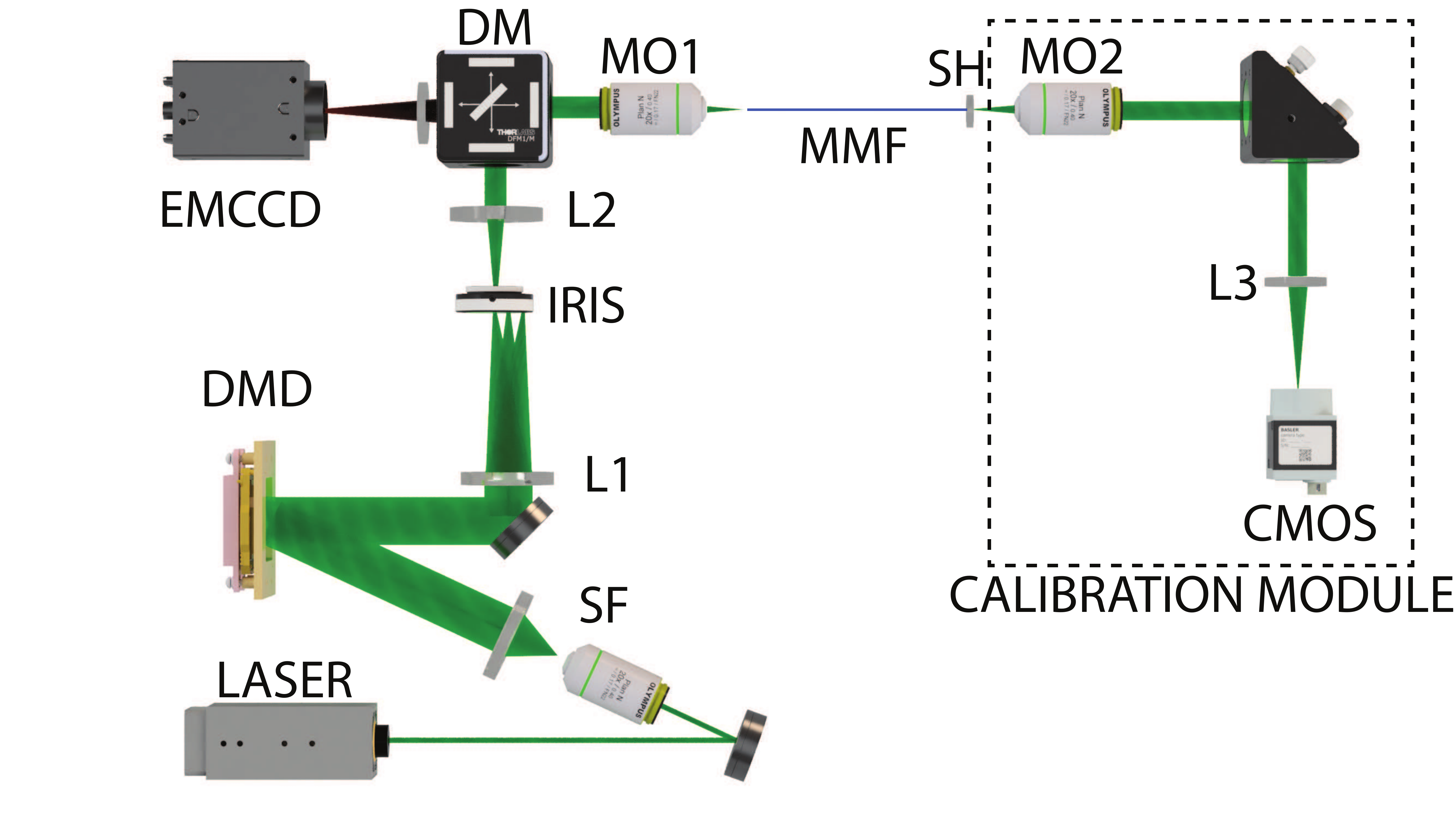}
\caption{Sketch of experiment setup for fluorescence imaging through a multimode fiber (MMF).}
\label{fig:BoulderSetup}
\end{figure}


\end{document}